\documentclass[twocolumn]{revtex4}
\usepackage{graphicx,mathtools,float}

\begin{document}

\title{\huge{A simple method to estimate fractal dimension of mountain surfaces}}
\author{{Kiran M. Kolwankar$^1$ and Nakul N. Karle$^2$} \\
$^1$Department of Physics, Ramniranjan Jhunjhunwala College, Ghatkopar(W), Mumbai 400 086\\
$^2$Department of Physics, University of Mumbai, Vidyanagari, Santacruz(E), Mumbai 400 098\\
E-mail: kiran.kolwankar@gmail.com, nakulkarle@gmail.com
}

\begin{abstract}

Fractal surfaces are ubiquitous in nature as well as in the sciences. The examples range from the cloud boundaries to the corroded surfaces. Fractal dimension gives a measure of the irregularity in the object under study. We present a simple method to estimate the fractal dimension of mountain surface. We propose to use easily available satellite images of lakes for this purpose. The fractal dimension of the boundary of a lake, which can be extracted using image analysis softwares, can be determined easily which gives the estimate of the fractal dimension of the mountain surface and hence a quantitative characterization of the irregularity of the topography of the mountain surface. This value will be useful in validating models of mountain formation

\end{abstract}

\maketitle

Fractals are irregular sets or objects whose, suitably defined, dimension usually has a non integer value. Though the concept was known to mathematicians much before it was Benoit Mandelbrot \cite{BB} who popularized it and demonstrated its immense applications to other branches of sciences \cite{KF} \cite{kf} \cite{JF}. An early well known example of an object whose fractal dimension was computed is the coastline of Britain. Since then numerous examples, not only in the field of geology \cite{DL} but in the whole of sciences \cite{MF}, have been found where approximation by a set of fractional dimension turns out to be more appropriate than by a regular set. The other examples of fractals include mountains, clouds, river networks \cite{DL}, trees \cite{MF}, turbulent velocity fields, large biological molecules, structure of lungs and veins \cite{GA}, etc.
A rigorous definition of fractional dimension was given by Hausdorff in 1919. But this definition is difficult to use in practice and several more practical, though sometimes less accurate, definitions came into existence. The most widely used among them is the box-dimension. In this definition the given set is covered with boxes of some size and the number of boxes needed to cover the set is counted. As the box size is reduced this number is expected to grow and the irregularity of the set would govern the growth of this number. The box dimension is the exponent in the power law growth of the number of boxes needed to cover the set. So mathematically put \cite{KF}, we have, for a set F
\begin{equation}
dim_B F = {\lim_{\delta \to 0}} - \dfrac{logN(\delta)}{log(\delta)}
\end{equation}
where $N({\delta})$is the number of boxes of size ${\delta}$ needed to cover the set. The limit in this definition is only a mathematical idealization and should be taken only when mathematical sets are being analyzed. Objects found in nature have a length scale below which it fails to be fractal and the definition cannot be applied below that scale. If the range of scale on which this power law holds is large enough then one can say that characterizing it as a fractal is a better approximation than a smooth model. In practice, if a log-log plot of $N({\delta})$ and ${\delta}$ yields a good straight line over a large range (over a decade) of scales then one calls the object as a fractal and the negative of the slope of this line gives the box dimension.
Fractal surfaces or interfaces are the objects of interest in several studies. For example, it could be a naturally occurring mountain or a surface of a porous rock. Corrosion also gives rise to very irregular surface which has been shown to be a fractal in some cases. Molecular aggregation leads to fractal surface which is growing with time. Thus it becomes necessary first to characterize such surfaces with proper fractal dimension and then to understand the origin of the irregularity leading to a theory explaining the value of fractional dimension. Also, there are problems of interest where one needs to study some phenomenon in the presence of a fractal structure or near a fractal interface. For example, it could be diffusion on fractals to understand a fluid flow in porous rocks or it could be Brownian motion near fractal interface as a model of chemical reaction taking place near in the presence of a catalyst with irregular surface. This forms the second class problems one needs to tackle which involve fractals.
	In this paper, we first briefly describe our previous works involving fractal interfaces. First one is a simple statistical model of etching which gives rise irregular interface which under certain conditions turns out to be fractal. Another work in which we have studied statistics of Brownian bridges, that is, loops formed by a random walker starting and ending on a surfaces which is taken to be a fractal in the study under consideration. This is useful in correct modeling of reaction rates for reactions taking place near a catalyst having fractal surface. Then we report a simple method to estimate the fractal dimension of mountain surface from that lake’s boundary obtained from readily available satellite images.
	A simple statistical model for etching for disordered solid was proposed \cite{KM}. In this model, a lattice is considered and each lattice site is assigned a random number between 0 and 1 which is considered to be its activation energy. The rate of dissolution of this site is given by the Arrhenius law. This lattice is in contact with a solvent which is assumed to be available abundantly and whose concentration is also assumed to be constant. There exists an interface between the solvent and the solid which consists of those lattice sites which are in contact with the solvent. A Monte Carlo evolution of the interface consists of dissolution of a randomly chosen site, sampled according to the rate of dissolution, on the interface. The etching of one site exposes one or more sites of the lattice and interface changes its shape. The Arrhenius law brings in the temperature dependence in the evolution and simulations are carried out for different temperatures. Some interesting results were reported. At low temperatures a very irregular interface develops which can be characterized as a fractal. At high temperatures, the interface is only mildly irregular. Another interesting aspect was that the overall rate of reaction depends on the percolation threshold, a property of the lattice. We also observed avalanches in the reaction, that is, the reaction progresses intermittently. Sometimes a very large number of sites were etched in a short time whereas at other times few sites were etched. This avalanche dynamics followed a power law behavior.
In another work we studied the statistics of first return of a random walker near a fractal interface \cite{PL}. This is useful when, for example, a chemical reaction takes place near a surface. When the molecule comes near the surface there is a probability that the reaction takes place. If it does not then the molecule can diffuse away and will possibly react when it returns to the surface and so on. As a result, the statistics of Brownian loops near a surface is important in deciding the overall chemical kinetics. This statistics is known to follow Cauchy distribution for smooth surface which asymptotically decays as $x^{-2}$ power law. We carried out extensive simulations and showed that this law is not valid near fractal surfaces. A general law with a new power law exponent in place of 2 which also depends on the dimension of the surface replaces the Cauchy’s distribution.

\begin{figure} [H]

\includegraphics[width=8.5cm]{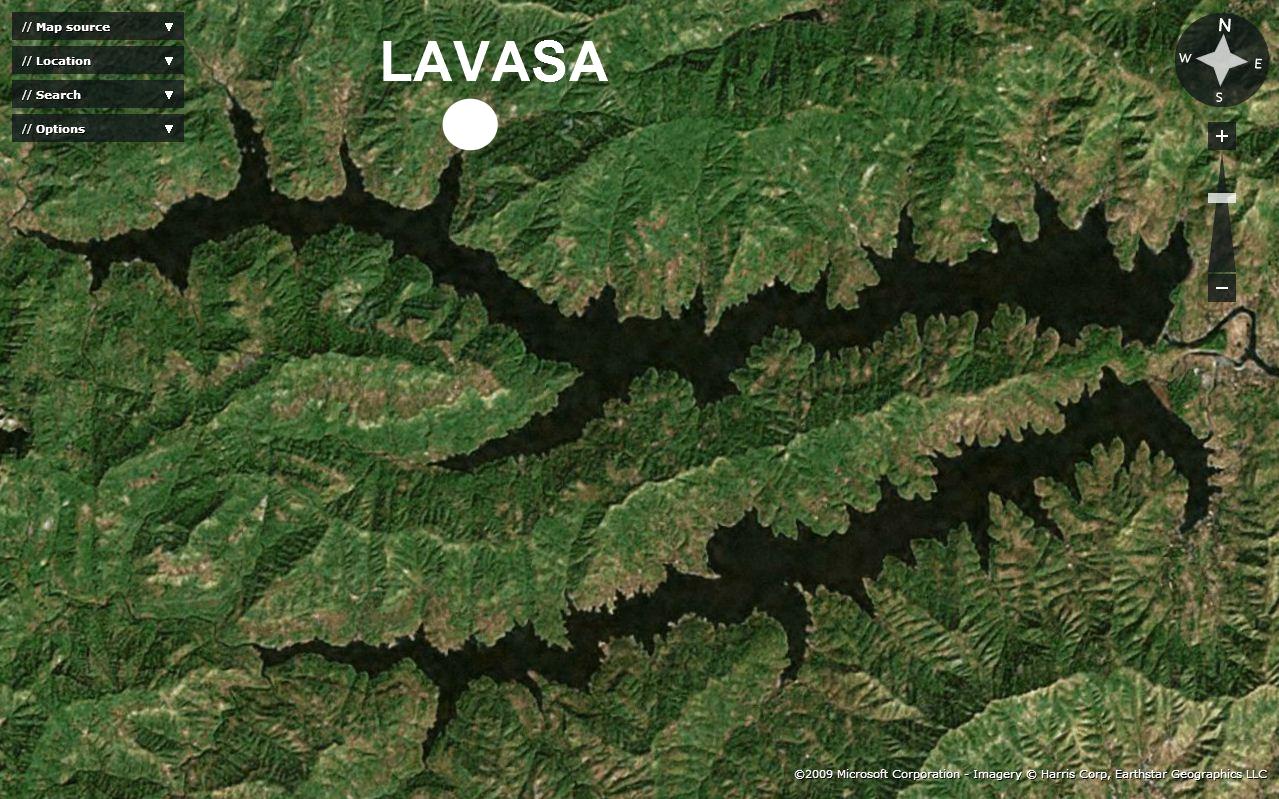}
\caption{Satellite image of two reservoirs Varasgaon (top) and Panshet (bottom) obtained from Bing. Lavasa is located on the banks of Varasgaon reservoir (indicated in the image by white spot) which has been analyzed here.}
\label{vr}

\end{figure}

Now we turn to the problem of estimation of fractal dimension of mountain surfaces using boundaries lakes or water reservoirs situated in the mountain. Satellite images of such water bodies are readily available. Also there exist numerous tools for image manipulation. It then is a simple matter to color the water body black and then extract the binary image then is a simple matter to color the water body black and then extract the binary image consisting only of the water body. After this, the edge detection algorithm is used to obtain the boundary of the lake.

\begin{figure} [H]
\includegraphics[width=8.5cm]{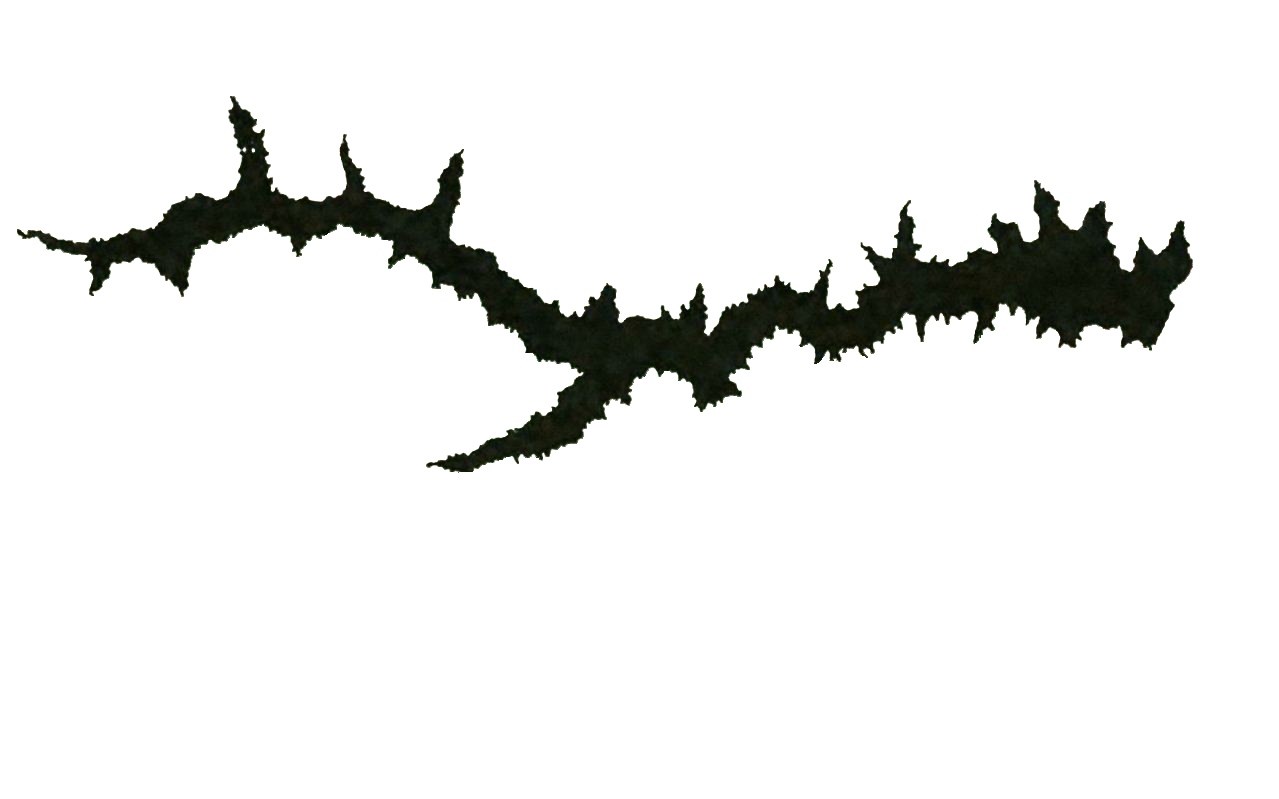}
\end{figure}
\begin{figure} [H]
\includegraphics[width=8.5cm]{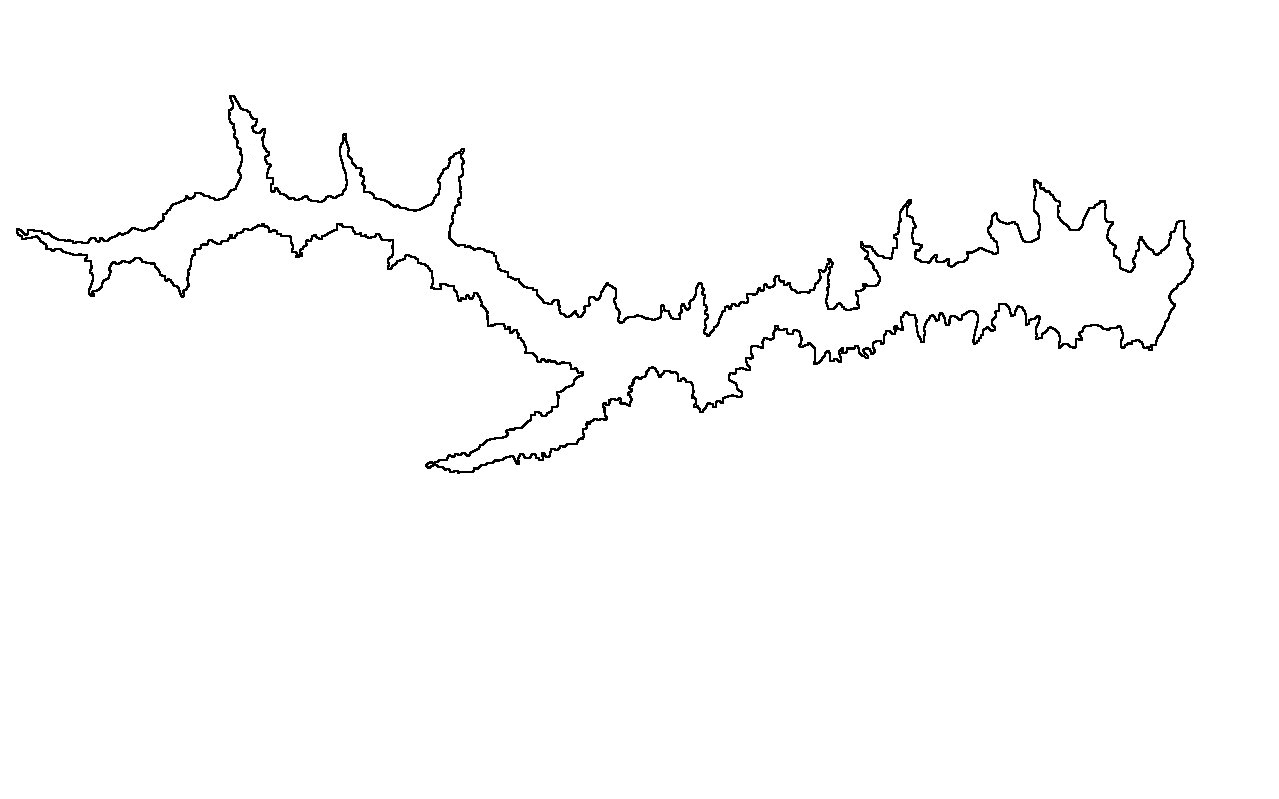}
\caption{The binary image of the Varasgaon lake after extracting the water body and then the boundary.}
\label{ol}
\end{figure}

Now the next step is to analyze the boundary again by using readily available fractal analysis tools. We have used ImageJ and Fractalyse softwares and we find the boundary to be fractal with a good power law fit from 40 meters to 2000 meters (Fig 3). The value of the slope turns out to be 1.32  which then is the dimension of the boundary. Clearly, in the case of reservoirs located in the mountains the irregularity in its boundary is a consequence of the irregularity of the mountains. Which, in turn, depends on the geological details of formation of that mountain. The boundary of a lake can be looked upon as a cross section of the mountain. Thus its dimension is the dimension of the cross section. Now if we assume that the dimension of such cross sections do not depend much on where the cross section is taken then the dimension of the lake’s boundary gives us an estimate of the dimension of the mountain. If $\alpha$ is the dimension of the boundary of a lake then 1+$\alpha$ is the dimension of the mountain.

\begin{figure}[H]
\includegraphics[width=9cm]{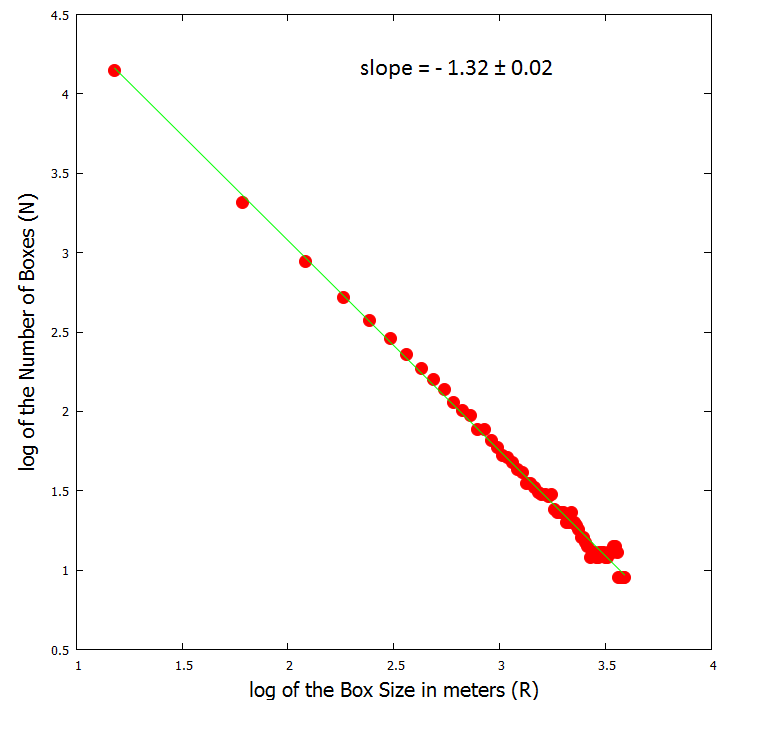}
\caption{Log-log plot of the number of boxes needed to cover the boundary with a given size of the box. The value of the slope and hence the dimension of the boundary is 1.32.}
\label{graph}
\end{figure}

There are several advantages of this simple method to estimate the dimension of mountain surface. Firstly, the satellite images of water reservoirs with sufficient resolution are now readily available as against the technological difficulty one would encounter in collecting the data regarding the height of the mountain at every point. Moreover, the information regarding the irregularity of a mountain as provided by lakes’ boundary is very local to that region. In other words, the dimension of lakes’ boundary can vary from place to place reflecting the changing irregularity of the mountains from place.
	The value of the dimension of the mountain thus obtained will be useful in validating models of mountains and their formation. As discussed by Mandelbrot in his book, the surface generated by a fractional Brownian process is taken as good mathematical model for mountain surface. Also, the physical models for formation of mountains would involve landslides, erosion etc. Constructing such models is challenging task. Sapoval et al. \cite{BS} recently proposed a model to understand the fractal dimension of coastlines. According to the authors, the final shape of the coastline is a result of self-organization between two competing forces, one is the erosion of the coast because of the bombarding waves and other is the damping of the wave by the irregular nature of the coast.
	To conclude, we have proposed a simple method to estimate the dimension of a mountain surface by measuring the dimension of a lake’s boundary which is located in the mountain. This makes use of readily available satellite images of water bodies located in the mountains. We have demonstrated the procedure on lake Varasgaon stituated in the western ghats near the western coast of India. The value turns out to be 1.32 and hence the dimension of the mountains around is approximately equal to 2.32. Clearly, this value will vary from place to place and will help us in validating the models of mountain surfaces and mountain formation.

\emph{Acknowledgement:} The satellite image used in this work was taken from Microsoft Bing website.

\end{document}